\documentclass[pra,showpacs,floatfix,superscriptaddress]{revtex4}

\usepackage{graphicx}
\usepackage{mathrsfs}
\usepackage{amsmath,amsfonts,amssymb}
\usepackage{subfigure,float}
\begin{document}
%\restylefloat{figure}

% title and related---------------------------------------------
\title{Critical examination of the inherent-structure-landscape
  analysis of two-state folding proteins} 

\author{Johannes-Geert Hagmann}
\affiliation{Universit\'e de Lyon; Ecole Normale
  Sup\'erieure de Lyon, Laboratoire de Physique,  CNRS, 46 All\'ee
  d'Italie, 69364 Lyon, France}  
\author{Naoko Nakagawa}
\affiliation{College of Science, Ibaraki University, Mito, Ibaraki 310-8512,
  Japan} 
\author{Michel Peyrard} 
\affiliation{Universit\'e de Lyon; Ecole Normale
  Sup\'erieure de Lyon, Laboratoire de Physique,  CNRS, 46 All\'ee
  d'Italie, 69364 Lyon, France}  
\date{\today}

\pacs{87.15.A-, 87.15Cc, 05.70.-a}

\begin{abstract}
Recent studies attracted the attention on the inherent structure landscape
 (ISL) approach as a reduced description of proteins allowing to map their
 full 
thermodynamic properties. However, the analysis has been so far limited to
a single topology of a two-state folding protein, and the simplifying 
assumptions of the method have not been examined. In this work, we construct
the thermodynamics of four two-state folding proteins of different sizes and
secondary structure by MD simulations using the ISL method, and critically 
examine possible limitations of the method. Our results show that the ISL 
approach correctly describes the thermodynamics function, such as the 
specific heat, on a qualitative level.  Using both analytical and 
numerical methods, we show that some quantitative limitations cannot be 
overcome with enhanced sampling or the inclusion of harmonic corrections.
\noindent
\end{abstract}

\maketitle

\section{Introduction}
The biological and physical properties of proteins are compelling for many
reasons. While just a small amount of the nowadays hundreds of thousands known
protein sequences are experimentally characterized, the variety of their
functions is overwhelming. Though the
structure has been resolved only for a subset of these sequences, 
the number of stable folds that are expressed in
nature is seemingly small compared to the number of sequences. The
relationship between fold and function is far from obvious, and examples such
as intrinsically unstructured proteins and multi-functional folds resist
simple schemes for classification. The question what really makes a protein
functional hence needs to be addressed in the context of its specific
biological environment.

From a physical point of view, an attempt to find
some unifying concepts for the interpretation of dynamics and thermodynamics
is the description of proteins in term of energy
landscapes\cite{frauenfelder}, in which the evolution of the system is related
to the dynamics on a high-dimensional rugged energy surface. The existence of
local minima, connected by saddles of different barrier height and rank, leads
to a distribution of timescales that are reflected in the dynamics of the
proteins. 

Although the energy landscape provides a reduced description, 
 the complex interactions in proteins and
their interaction with the environment, which involve multi-body
interactions and subtle effects of charges, make its complete characterization 
neither experimentally nor theoretically
conceivable. This situation is somewhat reminiscent of other complex systems
such as glasses which, though being more homogeneous systems, share the
property of displaying a high-dimensional landscape leading to complex
dynamics. The energy landscape picture is useful for a qualitative analysis of 
protein properties, but for quantitative studies, an exhaustive sampling and 
and the full characterization are practically infeasible for this 
high-dimensional representation. Therefore, in order to obtain comprehensive
quantitative predictions on generic protein properties from the information 
on the landscape, the picture needs to be simplified. 
A recent work\cite{nakagawa1} has shown that a reduced
description of the energy landscape, originally devised for the analysis of
super-cooled liquids by Stillinger and Weber\cite{stillinger}, can
successfully capture the essential thermodynamic aspects of folding in the
context of a simplified protein model. In particular, it was shown that the
density of states constructed from the local minima of the energy landscape,
called inherent structures, 
can be used to compute the most important thermodynamic observables. This
finding is important because it provides a general scheme for theoretical
studies of protein thermodynamics, showing how 
the relevant information can be quantitatively
accessed from its imprint on the potential energy surface. The approach has
consequently been used in the context of studies of the folding properties of
a $\beta$-barrel forming protein\cite{kim}, the construction of the free
energy landscape by mechanical unfolding\cite{imparato}, and the network of
native contacts\cite{wall}. Other earlier works including inherent structure
 analysis but not necessarily seeking to characterize the full 
thermodynamics of folding can be found in \cite{guo,shea}.

However the validity of an analysis based on the inherent structure landscape
(ISL) must be critically examined because the method involves a fundamental
assumption which could be questioned: The vibrational free energy within the
basin of attraction of an inherent structure is assumed to be independent of the
basin. A recent study \cite{wall} tried to go beyond this approximation by
assuming that the vibrational free energy can depend on the energy of the
inherent structures. 
Still, the question is subtle as we show
in the present work that, even when the vibrational free energy depends on the
inherent structure energy, the derivation of thermodynamic quantities such as
the specific heat from the ISL can be validly carried without any change in the
procedure. 
Therefore an understanding of the limits of the ISL approach requires a
deeper analysis. This is the aim of the present work.

We proceed in two steps. In a first step (Section \ref{section3}), 
after briefly summarizing the ISL
formalism and introducing the protein model used in
Section  \ref{section2}, we test the validity
of the ISL approach by comparing its results to the data obtained from equilibrium
molecular dynamics for a set of structures.
We selected four previously unstudied two-state folding
proteins of varying size and secondary structure elements. In a second step (Section ~ \ref{section4}), we
critically revisit the major hypotheses of the ISL approach, as well as its
practical limitations, such as the sampling of the phase space, and suggest 
routes for improvements. 
Finally, we
summarize our findings in Section \ref{section5} and give an outlook on possible
future studies that stem from our results.

\section{Methods}
\label{section2}

\subsection{Inherent structure analysis}
\label{ihstheory}
In this subsection, we briefly review the major results of
\cite{nakagawa1,nakagawa2} on obtaining reduced thermodynamics from an
analysis of the inherent structures.

The method is general and not bound to a specific protein model, provided
the phase space of the protein can be explored by molecular dynamics, and that the
energies of the visited states can be calculated.
From simulations at fixed temperature close to the
folding temperature $T_f$, which insures that the system evolves in a large part of the configuration space, the local potential minima, labelled by $\alpha_i$, 
are determined by
conjugate gradient minimizations performed 
at fixed frequency along a molecular dynamics trajectory. The
global minimum $\alpha_0$ is defined as the reference ground state with zero
energy. Let $\lbrace x_i \rbrace,\ i=1,2,...,3N$, denote the $3N$ Cartesian
coordinates of the $N$-particle system, and $V(\lbrace x_i\rbrace)$ its
potential energy function. The probability to find a particular minimum
$\alpha_i$ with potential energy $e_{\alpha_i}$ can be
written as 
\begin{eqnarray}
p(\alpha_i,T)&=&\frac{1}{Z(T)}\int_{B(\alpha_i)}d^{3N}x\ e^{-\beta V(\lbrace
  x_i \rbrace)}=\frac{1}{Z(T)}e^{-\beta e_{\alpha_i}}
\int_{B(\alpha_i)}
d^{3N}x\ e^{-\beta \Delta V_{\alpha_i} (\lbrace x_i \rbrace ) } \ \ \ , \label{eq1}  
\end{eqnarray}
where $\Delta V_{\alpha_i} = V - e_{\alpha_i}$ , $Z$ denotes the
configurational part of the partition function and $B(\alpha_i)$ is the basin
of attraction of the minimum $\alpha_i$.
With the definition\begin{eqnarray}
e^{-\beta F_v(\alpha_i,T)}&:=&\int_{B(\alpha_i)} }d^{3N}x\ e^{-\beta \Delta
    V_{\alpha_i}(\lbrace x_i \rbrace) \label{fdefine} \ \ \ , 
\end{eqnarray}
the unknown integral over the complex landscape of the basin of attraction
$B(\alpha_i)$ is summed in an free-energy like
function $F_v(\alpha_i,T)$ which in principle depends both on the nature of
the basin and temperature. Notice that although we use the index $v$ like
''vibrational'', $F_v(\alpha_i,T)$  is obtained 
from the full nonlinear integral over $B(\alpha_i)$, 
and not from its harmonic approximation.\\The inherent structure landscape approach makes two
key assumptions\cite{nakagawa1} which enable to considerably reduce the amount
of information needed on the landscape while keeping its most important
features. 
\begin{itemize}
\item{(A1) The function $F_v(\alpha,T)$ for two minima $\alpha_1,\alpha_2$
  that are distinct but close in energy, $e_{\alpha_1} \approx e_{\alpha_2}$,
  is the same for both minima:  $F_v(\alpha_1,T)\approx
  F_v(\alpha_2,T)$. Consequently, $F_v(\alpha,T)\approx F_v(e_{\alpha},T)$.} 
\item{(A2) The function
$F_v(e_{\alpha},T)$ does not vary significantly for different minima,
  i.e. $F_v(e_{\alpha},T)\approx F_v(T)$.} 
\end{itemize}
Both assumptions were discussed in \cite{nakagawa2}. In section \ref{section4},
we show that assumption (A2) can actually be relaxed to the less strong form
$\beta F_v(e_{\alpha},T)\approx f_{v}(e_{\alpha})+ \beta F_{v}(0,T)$
while most calculations remain feasible and some of the thermodynamic
variables unchanged. With these assumptions, the contribution from the
function $F_v$ factorizes in the numerator and denominator of 
 (\ref{eq1}) so that it can be eliminated to
give 
\begin{eqnarray}
\label{eq:pz}
p(\alpha_i,T)&=&\frac{1}{Z_{IS}(T)}e^{-\beta e_{\alpha_i}} \ \ \ \ , \ \ \ \ 
Z_{IS}=\sum_{\alpha=\alpha_0}^{\alpha_{max}}e^{-\beta e_{\alpha}}\ \ \ ,
\end{eqnarray}
where the sum in the partition function includes all inherent structures found
from the global minimum $\alpha_0$ to the minimum $\alpha_{max}$ having 
the highest energy. Here, the energy scale is shifted such that the energy of the global minimum $\alpha_0$ is zero.  
Introducing an energy density function for the inherent structures
$\Omega_{IS}(e)=\sum_{\alpha=\alpha_0}^{\alpha_{max}}\delta(e-e_{\alpha})$,
the probability to find a minimum in the interval
$[e_{\alpha},e_{\alpha}+de_{\alpha}]$ at temperature $T$ is 
\begin{eqnarray}
\label{eq:piszis}
P_{IS}(e_{\alpha},T)de_{\alpha}&=&\frac{1}{Z_{IS}}\Omega_{IS}(e_{\alpha})e^{-\beta
  e_{\alpha}}\ de_{\alpha} \ \ \ \ , \ \ \ \ Z_{IS}=\int_{e_{\alpha_0}}^{e_{\alpha_{max}}}
de_{\alpha}\  \Omega_{IS}(e_{\alpha}) e^{-\beta e_{\alpha} } \ \ \ .
\end{eqnarray}
For the model used in this work, the low energy minima are in
practice sparsely separated in energy. As the ground state is isolated, one
obtains $Z_{IS}(T)=1/p_0(T)$ with the probability of the ground state $p_0(T)$, so that the inherent structure density of
states can be estimated from the probability to be in the basin of attraction of a minimum in a fixed temperature simulation at temperature $T_{MD}$ as  
\begin{eqnarray}
\Omega_{IS}(e_{\alpha})&=&\frac{e^{\beta_{MD}
    e_{\alpha}}}{p_0(T_{MD})}P_{IS}(e_{\alpha},T_{MD})\ \ \ . \label{ihs} 
\end{eqnarray}
Though above we have chosen a continuous notation to simplify the 
equations, it should be
noted that the density built from an estimate of the probability density function
of sampled minima for the present model in practice always comprises discrete and 
continuous parts which can be integrated separately.
Once the inherent structure density of states is known, one can compute the
inherent structure partition function $ Z_{IS} $ from (\ref{eq:piszis}),
from which all thermodynamics functions can be derived, 
including the free energy $F_{IS}$ and the internal
energy $U_{IS}$. Given that most states of the system
can be sampled close to the folding temperature, it is sufficient to simulate
the system at a single temperature $T_{MD}$ to construct the inherent structure
landscape, in contrast to the full thermodynamics where one needs to sample
different ranges of temperatures. Therefore, the ISL approach can be
computationally very efficient. In the following, we will restrict ourselves to the
computation of specific heat $C_{V,IS}$ being a quantity of
fundamental importance in a physical system, as it is sensitive to 
fluctuations and, for instance, shows a clear signature of phase
transitions. It can be deduced from numerical
derivatives of the partition function $Z_{IS}$ through 
\begin{eqnarray}
 C_{V}=T\left(\frac{\partial S}{\partial T}\right)_V \ \ \ ,
\end{eqnarray}
and hence
\begin{eqnarray}
  C_{V,IS}=T\left(\frac{\partial^2 (\beta^{-1}\log(Z_{IS})}{\partial T^2}
  \right)_V  \ \ \ \ \ . 
\end{eqnarray}
\subsection{Model and selected proteins}

Since our goal is to analyze the validity of the ISL approach and not to
derive quantitative data for a particular protein, we decided to choose a
simplified model, which allows the sampling of phase space at a
reasonable computational cost. However the model must be rich enough to
properly describe the complex features of its physics, and should be able to
distinguish between proteins which differ, for instance,
in their secondary structure. 
We use frustrated off-lattice G\=o-models identical to the ones introduced in
\cite{nakagawa1} because they provide 
a good compromise between all-atom simulations
 and simplified models that do not fully describe the
geometry of a protein.
These models provide a representation with a single particle per
residue centered at the location of each $C_{\alpha}$-atom. For details on the
model and the parameters, we refer to \cite{nakagawa1,nakagawa2} and a brief 
review in the appendix \ref{ap1}. 
Although
the validity of such models to provide a faithful representation of protein
folding is a recurrent subject of debate, off-lattice G\=o-models have been
successfully 
used to study folding kinetics \cite{karanicolas}  and the mechanical
resistance of proteins \cite{paci}. From a physical point of the view, despite
a strong bias towards the ground state, these models have a complex energy
landscape with a large number of local minima well suited for the analysis in
terms of inherent structures.\\As 
the results of the ISL approach depend on the density of states of
the inherent structures, for a reliable test of the method it is important to
examine examples which could differ in their properties, i.e. to investigate 
proteins of different size and structure.
To test the inherent structure approach beyond the previously employed
immunoglobulin (IG) binding domain of protein G (2GB1), we selected four two-state
folding proteins of varying size and folds from the PDB database\cite{pdb}:
the trp-cage mini-protein construct (1L2Y, 20 residues, $\alpha$-helical), the
ww domain FPB28 (1E0L, 37 residues, $\beta$-sheets), the src-SH3 domain (1SRL,
56 residues, $\beta$-sheets) and ubiquitin (1UBQ, 76 residues,
$\alpha-\beta$-fold). The motivation for these choices is discussed
in Section \ref{section3} for each protein (see insets of figures \ref{fig1}-\ref{fig4} for
structural representations of these four proteins drawn with pyMol\cite{pyMol}).
 The positions of the
$C\alpha$-atoms of the PDB files is chosen as a reference for the construction
of the G\=o-model. In the case of NMR resolved structures, the first structure
is selected as the reference. The native contacts of the model were
established according to the distances between atoms belonging to different residues. A native
contact is formed if the shortest distance belonging to atoms of two different
residues is smaller than $5.5\AA$. The number of native contacts according to this criterion are:
$N_{nat}=91$ for the ww domain, $N_{nat}=225$ for ubiquitin, $N_{nat}=216$ for
src-SH3 and $N_{nat}=36$ for trp-cage. This definition is simple and
 includes some arbitrariness. There exist other methods for probing
contacts between side-chains, e.g. by invoking the van der Waals radii of
residue atoms and solvent molecules\cite{sobolev}. Though using the latter
method preserves the main structure of the contact map, it leads to
quantitative differences in the the number and location of contacts along the
sequence. Consequently, one can expect that the topology of the energy surface
and key thermodynamic properties such as the folding temperature are also
altered when the definition of the contact map is varied. For the purpose of the
present study which does not attempt to give a quantitative description of
side chain contacts, and focusses on global properties of the landscape
rather than its detailed relation to the network of contacts, the cutoff-based
approach is acceptable.\\Molecular dynamics simulations were performed using the
Brooks-Br\"unger-Karplus algorithm \cite{bbk} with a time-step of $dt=0.1$ and
a friction constant of $\gamma=0.01,0.025$ (all units in this section are
dimensionless, see \cite{nakagawa1} for details). To ensure equilibration, the
system was thermalized starting from the native state (PDB coordinates) for
$t=2\cdot 10^5$. The simulation time for a single temperature point and a
single initial condition was $t=2\cdot 10^7$, and the data obtained for both
inherent structure sampling (fixed temperature) and thermodynamic sampling
(variable temperature) were averaged over various initial sets of
velocities. Minimization was performed using the conjugate gradient method
with the Polak-Ribi\`ere algorithm. To estimate the vibrational free energy at
the minimum, mass weighted normal mode analysis was performed using LAPACK
diagonalization routines. The second-order derivatives of the potential energy
function at the minimum were calculated by numerical differentiation of the
analytical first-order derivatives.

\section{Reduced and full thermodynamics of a set of model proteins}
\label{section3}
In this section, the validity of the ISL approach is tested 
by comparing the equilibrium
thermodynamics deduced from molecular dynamics simulations
to the reduced thermodynamics from inherent structure sampling. As
discussed in Section \ref{section2}, we evaluate the
specific heat $C_V$ as a function of temperature
as a representative example of the thermodynamic observables.

\subsection{src-SH3}

\begin{figure}
\centering
\includegraphics[width=6.6cm]{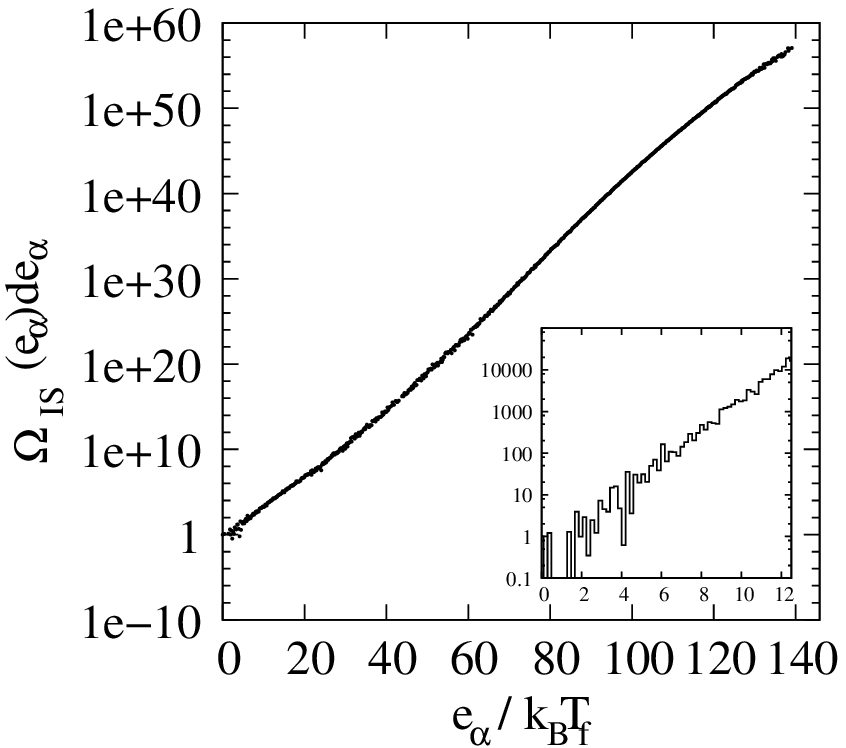}
\includegraphics[width=9.2cm]{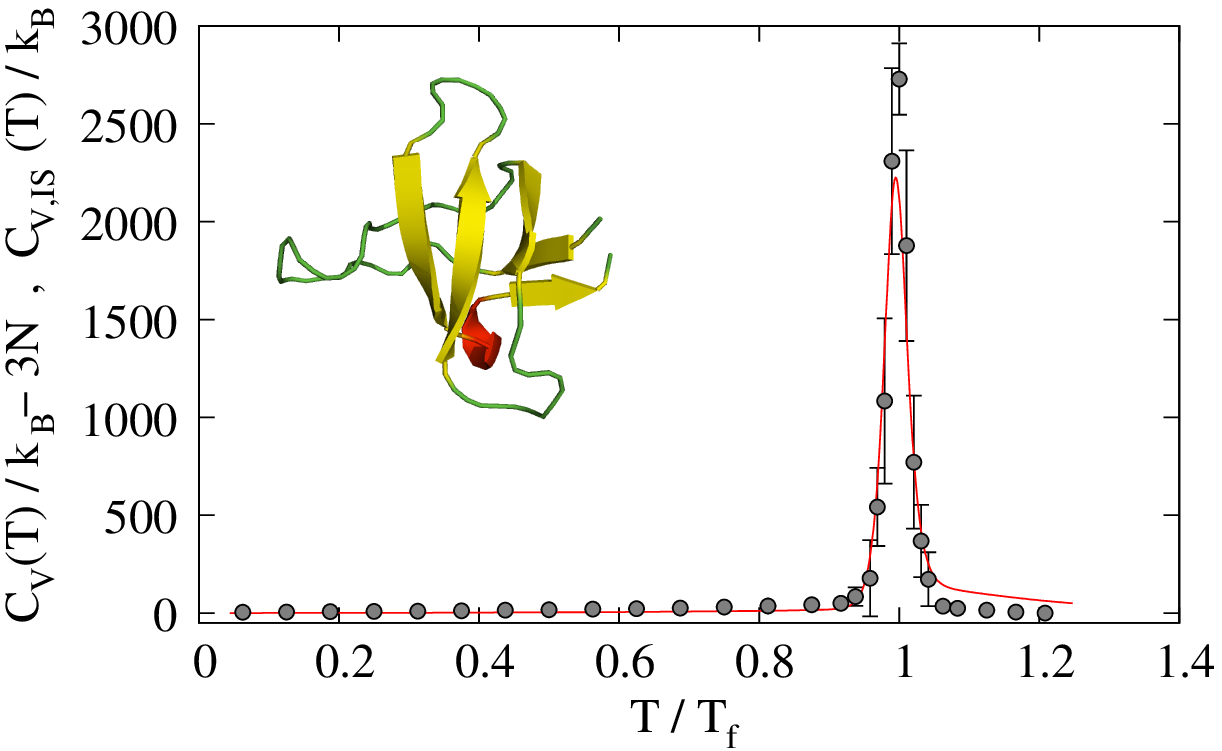}
\caption{Results for src-SH3. \textit{Left:} Inherent structure density of states
  $\Omega_{IS}(e_{\alpha})$. The inset shows a close up on the low-energy
  range. The size of the energy bins for the density estimate is $\Delta\ E=0.2\ k_BT_f$. \textit{Right:} Comparison of the specific heat from equilibrium
  trajectories $C_V(T)$ (points), from which the specific heat of a harmonic
  oscillator in $3N$ dimensions has been subtracted, and $C_{V,IS}(T)$ from
  inherent structure analysis (solid red line); see text for further
  explanations.} 
\label{fig1}
\end{figure}

The src-SH3 domain was chosen since it has the same number of residues
as IG binding domain of protein G studied in
\cite{nakagawa1,nakagawa2}, but contrasts to the latter in terms of
structure. The src-SH3 domain is mostly composed
of $\beta$-sheets and does not contain an $\alpha$-helical-secondary structure
element (only five residues form a small right-handed helix segment). The
inherent structure density of states, shown on the left-hand side of
figure \ref{fig1}, was obtained from various simulations close to the folding
temperature ($T_{MD}\approx T_f$) and built from $\approx 72000$ minima according to
 (\ref{ihs}). After computing an energy histogram using $1000$ bins to yield an 
estimate of the inherent
structure probability density, energy bins with only a single count have
been discarded from the analysis to avoid a bias that could be introduced
by insufficiently sampled isolated minima. The right
hand side of figure \ref{fig1} shows a comparison between the temperature
dependence of the specific heat calculated from inherent structures,
$C_{V,IS}(T)$, and the temperature dependence of the specific heat calculated
from equilibrium molecular dynamics simulations at variable temperature
$C_V(T)$. The equilibrium thermodynamics has been determined by averaging the
results of 10 initial conditions per temperature step except 
for points close to
the transition where the results of 20 initial conditions have been used. Despite this
averaging, the variance, indicated by error-bars on the y-axis, is large in
the vicinity of the folding transition as the waiting time for a transition
to occur becomes comparable to the simulation time. For a harmonic system,
$C_V(T)=N_{dof}k_B/2$ where $N_{dof}$ is the number of degrees of freedom. At
low temperatures $T\ll T_f$, harmonic contributions are dominating, and the
difference between $C_V(T)$ and $C_{V,IS}(T)$ is approximately $3Nk_B$, which
is subtracted from $C_V(T)$ in figure \ref{fig1}.

Figure \ref{fig1} shows that the ISL approach is able to capture the main
features of the thermodynamics of the G\=o-model of the src-SH3 domain. 
The value of the folding temperature is correctly determined by the ISL approach,
but $C_{V,IS}$ underestimates the maximum by more than $20\%$ if the highest
point of $C_V(T)$ is selected as a reference. This discrepancy at the maximum
had previously also been observed for the inherent structure analysis of the
IG binding domain of protein G \cite{nakagawa1}. On the other hand, towards
higher temperatures, $C_{V,IS}(T)$ decays slower than $C_{V}(T)$. 

\subsection{ubiquitin}

\begin{figure}
\centering
\includegraphics[width=6.6cm]{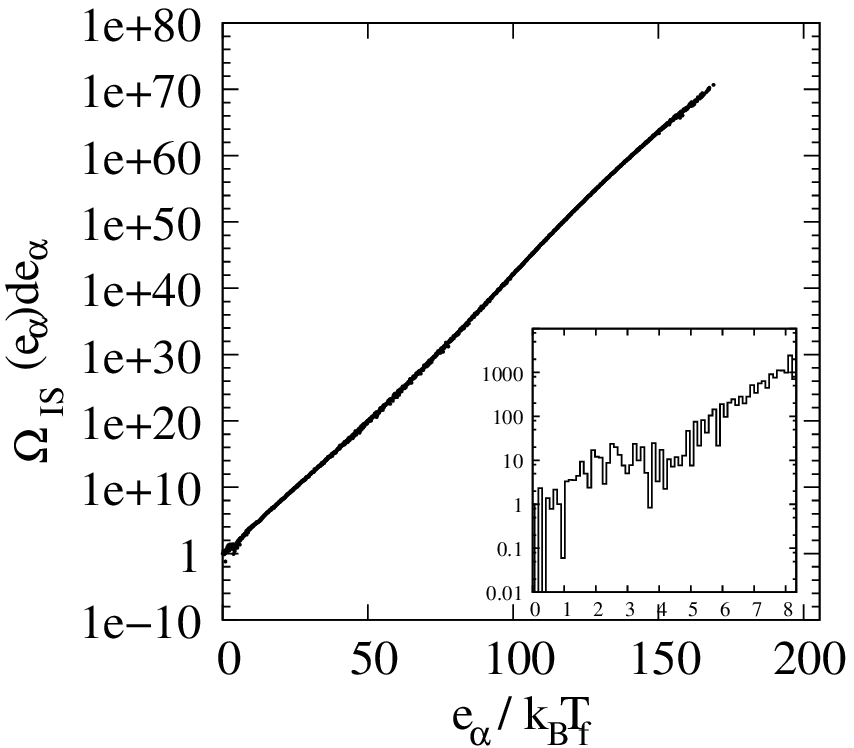}
\includegraphics[width=9.2cm]{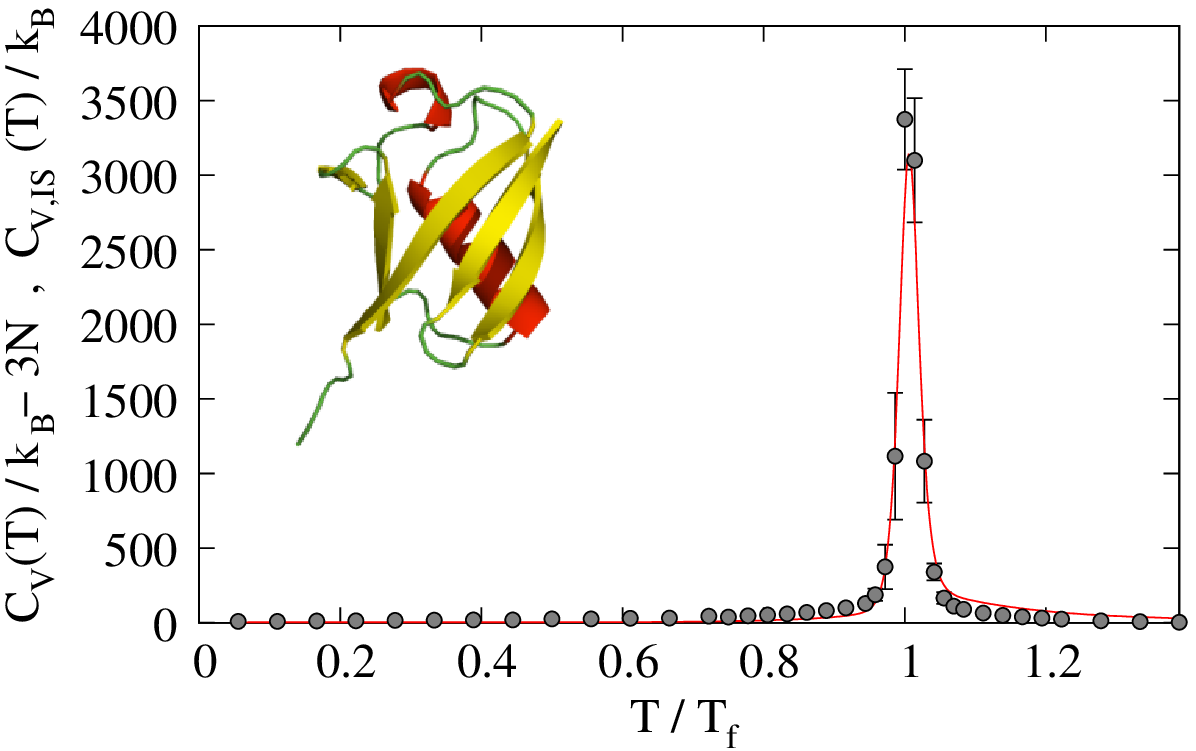}
\caption{Results for ubiquitin, see caption of figure \ref{fig1} for annotation.The size of the energy bins for the density estimate is $\Delta\ E=0.24 \ k_BT_f$. \hspace{4in}}
\label{fig2}
\end{figure}
Ubiquitin, with its 76 amino-acids, 
is a protein with a fairly rich secondary structure
since it contains a $\alpha$ helix and 5 $\beta$ sheets.
Similarly to the src-SH3 domain, the ubiquitin G\=o-model 
presents a sharp folding
transition associated with a large peak in the specific heat (see right-hand side of
figure  \ref{fig2}). The specific heat
$C_V(T)$ was estimated from averages on 8 initial conditions
per temperature step, and $\Omega_{IS}(e_{\alpha})$ was obtained using $\approx
79000$ minima from several independent trajectories close to the folding
temperature using a histogram of $2000$ bins. The agreement between the full thermodynamics 
and the ISL approach is better than for src-SH3,
though similar trends of discrepancies can be observed.

\subsection{ww domain}

\begin{figure}
\centering
\includegraphics[width=6.4cm]{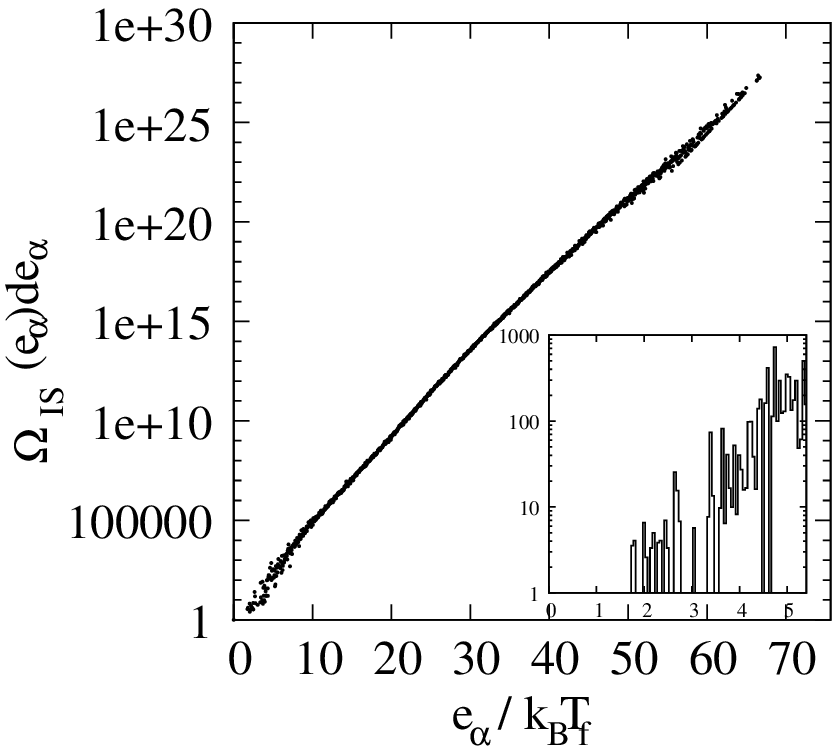}
\includegraphics[width=9.4cm]{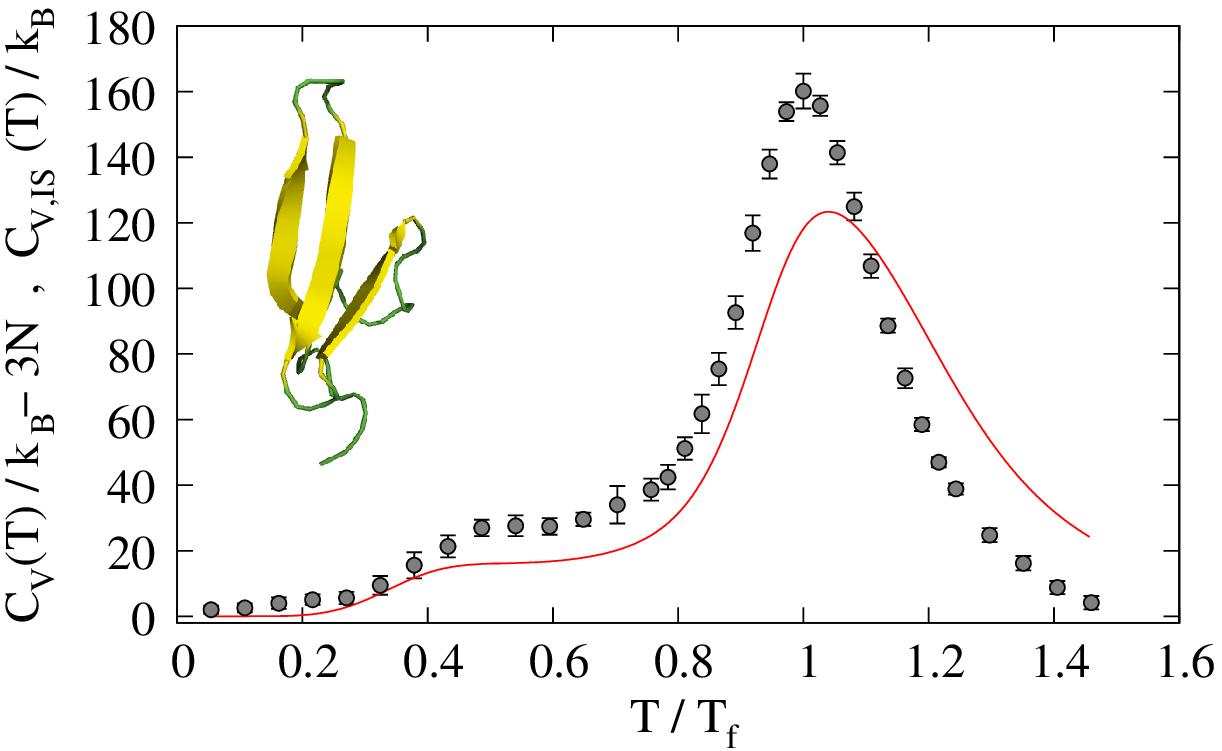}
\caption{Results for the ww domain, see caption of figure \ref{fig1} for
  annotation. The size of the energy bins for the density estimate is
  $\Delta\ E=0.05 \ k_BT_f$.\hspace{4in}}  
\label{fig3}
\end{figure}
To contrast with sharp two-state transitions of protein G (56 residues),
src-SH3 (56 residues) and ubiquitin (76 residues), we selected smaller
structures, the ww domain (37 residues) and the trp-cage (20 residues), 
to examine 
the performance of the inherent structure approach for less
structured proteins, showing a broader transition. For such small
protein domains, the validity of the G\=o-model can be questioned as the model 
is built from the geometrical structure of the folded state. For 
small molecules the discrimination between folded and unfolded states becomes
subtle due to fluctuations covering a large part of the accessible
configurational space  
in a broad range of temperatures. The 
point in selecting these structures is not to asses the validity of
the G\=o-model itself, but to test the 
ISL approach in very stringent cases to highlight possible
limitations. The density of states $\Omega_{IS}(e_{\alpha})$ was
obtained from $\approx 
79000$ minima from several independent trajectories slightly above the folding
transition (see figure  \ref{fig3}) using a histogram of $2000$
bins. In contrast to the two previous cases, the histogram of minima
does not show a clear separation of basins of local minima associated
to the folded/unfolded state. We observe a difference in the apparent
shape of the density of states (left-hand side of figure  \ref{fig3}),
which is globally concave in contrast to the convex densities obtained
for src-SH3 and ubiquitin. The same shape was also found for protein G
in \cite{nakagawa1}, for which the relation between the concave shape
and the two-hump structure of $P_{IS}(e_{\alpha},T)$ was discussed in
the vicinity of the folding temperature. Moreover, by comparing the
insets of the left-hand side of 
figs. \ref{fig1}-\ref{fig3}, one observes that the low energy range of
$\Omega_{IS}(e_{\alpha})$ is more discrete, and states tend to lie less
densely packed.\\ 
The temperature dependence of $C_V(T)$ was obtained by averaging over 12
initial conditions. An interesting feature of the curve is the shoulder in the
low temperature range which indicates a partially unfolded structure
associated to the breaking of a small number of contacts. Comparing the
results of $C_V(T)$ and $C_{V,IS}(T)$, it is apparent that though the
specific heat reconstructed from inherent structure thermodynamics 
correctly captures
 the global shape of $C_V(T)$, including the existence of the
shoulder,  important deviations can be observed. Similarly to the
cases analyzed above, $C_{V,IS}(T)$ underestimates $C_{V}(T)$ at lower
temperatures while giving an overestimation at high temperatures. In contrast
to the results for larger proteins, we also observe a significant shift of the
transition temperature. 

\subsection{trp-cage}

\begin{figure}
\centering
\includegraphics[width=6.4cm]{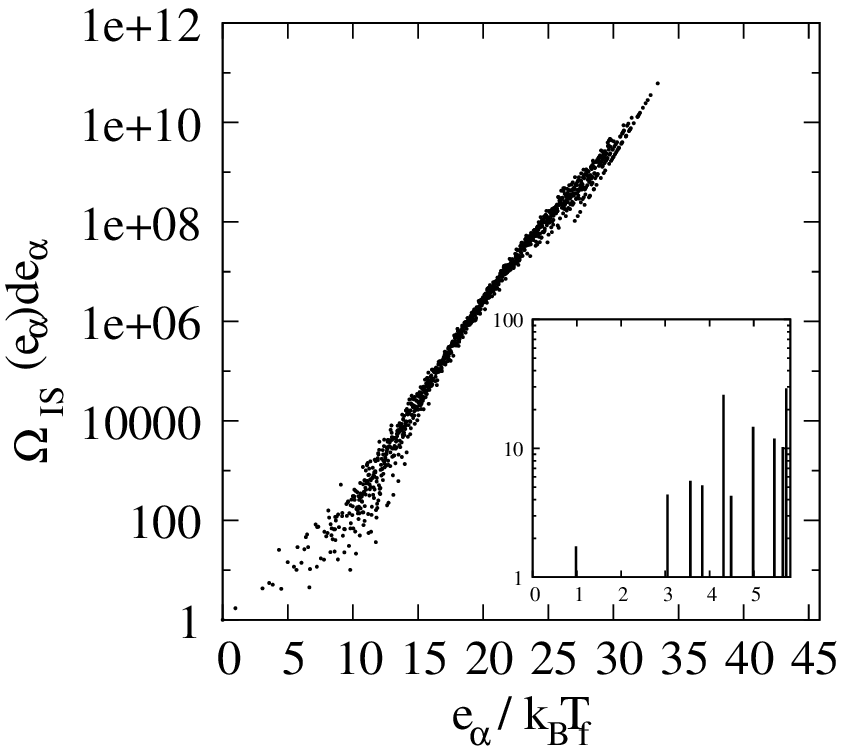}
\includegraphics[width=9.4cm]{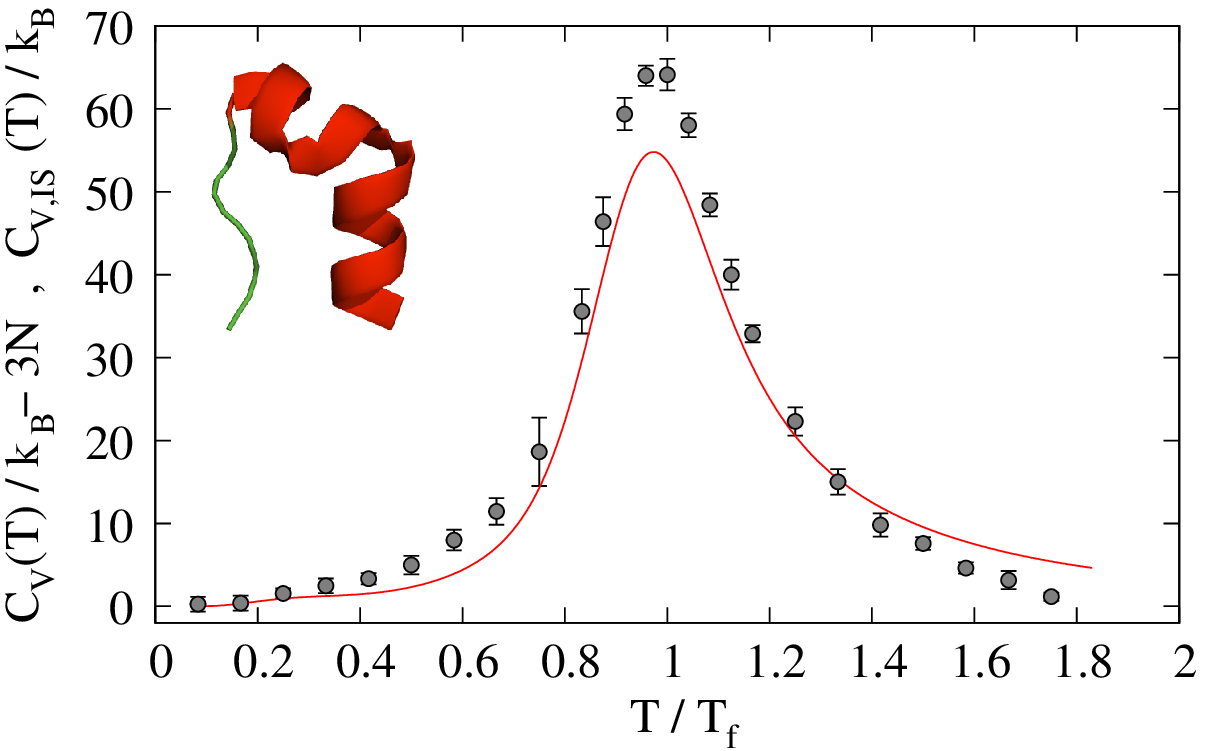}
\caption{Results for trp-cage, see caption of figure \ref{fig1} for
  annotation. The size of the energy bins for the density estimate is
  $\Delta\ E=0.02 \ k_BT_f$. \hspace{4in}} 
\label{fig4}
\end{figure}
With only 20 residues, the last protein fragment studied in this series is
also the smallest, and mainly consists of a single $\alpha$-helix. Its
inherent structure density of states $\Omega_{IS}(e_{\alpha})$ was estimated
from $\approx 90000$ minima sampled from several independent trajectories
close to the folding temperature using a histogram of $2000$ bins. For
such a small system, the low lying 
energy states are largely separated from each other, and the resulting density
of states presents large gaps in a relatively broad range of energies. The
continuum representation assumed for the inherent structure landscape is
certainly questionable in such a case. The equilibrium thermodynamics
were constructed from averages on 10 runs over different initial conditions. 
 It is
interesting to notice that the value of the specific heat computed with
reduced thermodynamics is still fairly close to the actual specific heat,
although we notice again that the peak is underestimated and a
high temperature tail is observed as in the previous cases. In contrast to
 the results for the ww domain, the folding temperature of the trp-cage protein
domain is correctly found by the analysis of inherent structures.

\subsection{Discussion}

\begin{figure}
\centering
\includegraphics[width=7.8cm]{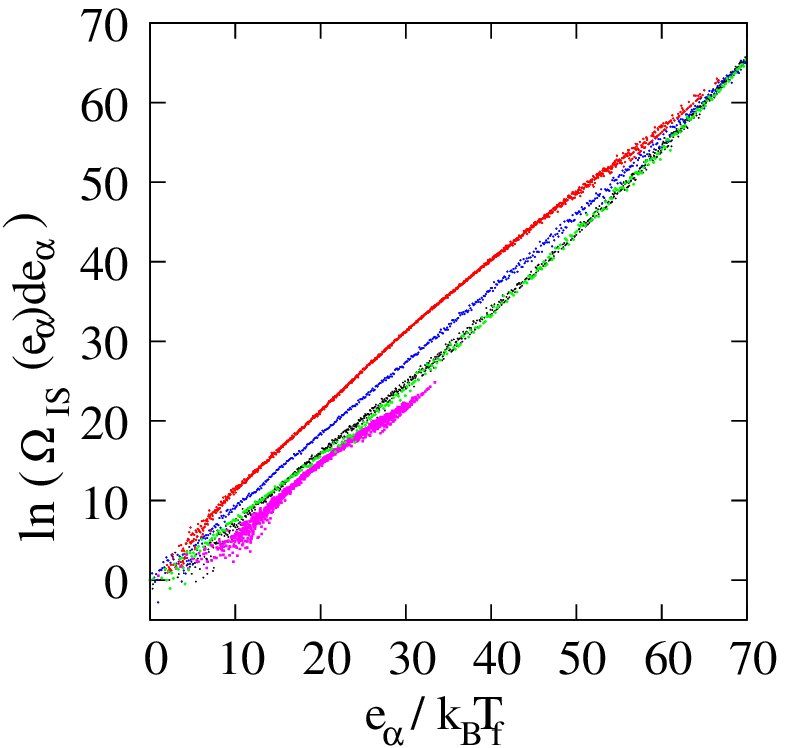}
\includegraphics[width=8.0cm]{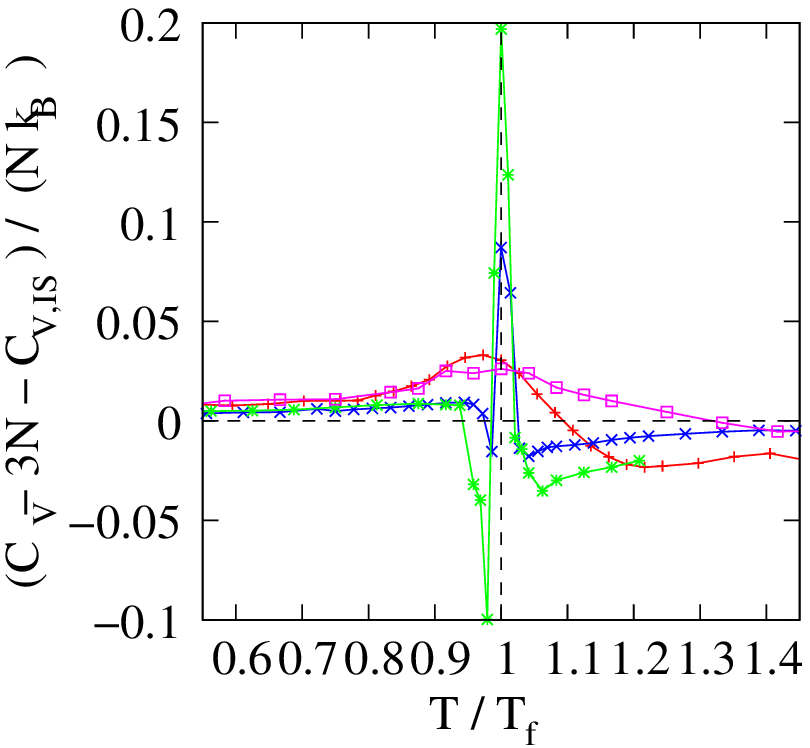}
\caption{\textit{Left:} Comparison between the different inherent structure
  density of states $\Omega_{IS}(e_{\alpha})$ including the data for the IG
  binding domain of protein G from \cite{nakagawa1}; the color coding is:
  protein G (black), src-SH3 (green), ww domain (red), ubiquitin (blue),
  trp-cage (magenta). \textit{Right:} Deviation between the specific heat
  obtained from full thermodynamics and inherent structure analysis.
} 
\label{newfig}
\end{figure}

Our studies of four proteins, combined with the earlier results on protein G
\cite{nakagawa1,nakagawa2}, allow us to describe some trends in the
inherent structure  
analysis of G\=o-model proteins.\\
For the density of states given by (\ref{ihs}), a general exponential
dependence,
$\Omega_{IS}(e_{\alpha}) \propto \exp(-e_{\alpha}/k_B T_0)$  
is observed for all proteins, with slightly different slopes for the
low energy states, corresponding to states occupied in the folded
configuration, and for the high energy states, occupied in the unfolded
configuration. The value of $T_0$ associated to the low energy range is
a good estimate of the folding temperature, as previously reported for
the case of 
protein G \cite{nakagawa1}.
Figure \ref{newfig} (left-hand panel), which compares the density of
states of the inherent 
structures for the four proteins shows that, when being presented
 in reduced units as a function of $e_{\alpha}/k_B T_f$, the functional form
of these densities is highly similar. 
For the large proteins that we studied,
 src-SH3 and ubiquitin (as well as protein G), the slope
is slightly larger in the high energy range than in the low energy range. The
converse is true for the small protein domains ww and trp-cage.
A formal calculation of the reduced specific heat $C_{V,IS}$ from a
bi-exponential density of inherent structure energies shows that this property
is related to the sharpness of the folding transition. A density of states
that is curved downwards for the energies associated to unfolded
configurations leads to the broad folding transition expected for small
protein domains.

The calculation of the specific heat $C_{V,IS}$ shows that the ISL approach is
able to determine the specific heat of a protein with reasonable accuracy, 
including the overall shape of the
folding transition. To obtain such an agreement the ground state probability
$p_0(T_{MD})$ must be sufficiently well sampled to ensure that the density of
inherent states is correctly normalized. 

However, our studies of several proteins shows that limitations exist since
systematic deviations from the full thermodynamics are apparent. At low
temperatures, and up to the transition temperature, $C_{V,IS}(T)$
underestimates the specific heat. The peak of $C_{V,IS}(T)$ is less
pronounced than expected from the equilibrium trajectories, and tends to
broaden towards higher temperatures ($T>T_f$) where $C_{V,IS}$ is larger than
$C_V(T)$. These deviation are shown in a comparative illustration in the right-hand side panel of figure \ref{newfig} for the four proteins. Our data
 also reveals that the G\=o model, with its strong bias towards the
 native state, does not yield qualitatively differences depending on 
the secondary structure of the protein under consideration. 
\\Owing to the results found for various proteins which show systematic
deviations from the results of equilibrium thermodynamics, it is important to
examine the assumptions made in approximating the inherent structure of
states, which we do in the following section.

\section{The limitations of the ISL approach}
\label{section4}
\subsection{Local normal mode analysis}
\label{NMA}

In Section \ref{section2} we introduced 
the major assumptions (A1) and (A2) of the ISL approach. The derivation
of thermodynamic quantities such as the specific heat is carried out as
if the free energy contribution within a basin of
attraction, defined by (\ref{fdefine}), did not depend on the particular
inherent structure $\alpha_i$. It is difficult to test this assumption
as it would in principle require the
determination of the complete basin on the energy landscape, including
the calculation of all the saddle points that determine the frontier of the
basin as well as the shape of the basin within this frontier. 
Still, one can at least compute $F_v(\alpha)$ {\em in the harmonic
  approximation,} as done also in \cite{wall}.

\begin{figure}
\centering
\includegraphics[width=8.0cm]{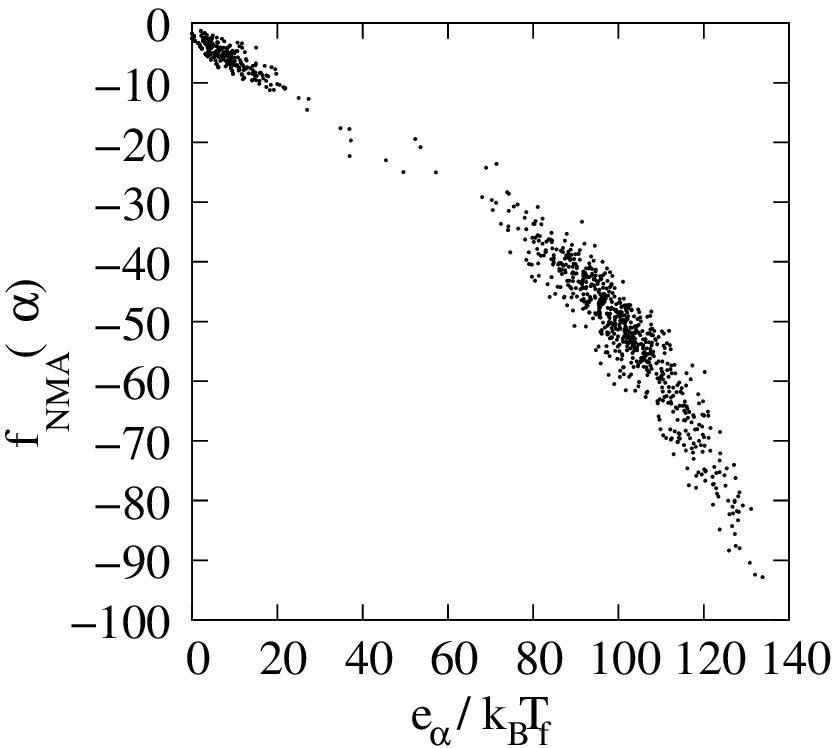}
\includegraphics[width=7.8cm]{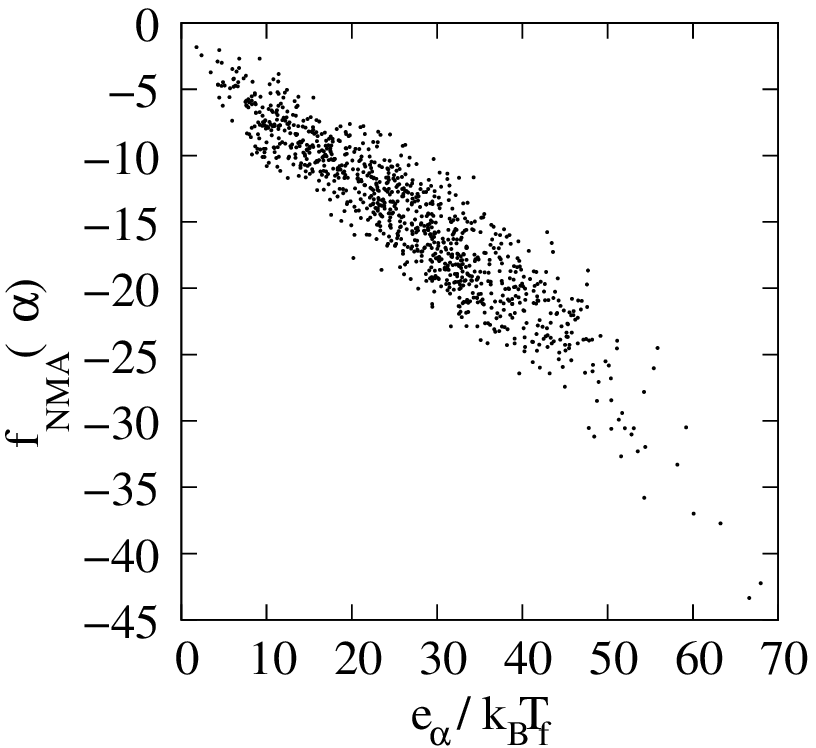}
\caption{The frequency dependent component $\sum_{q=1}^{3N-6} \log(\omega_q)$ of the vibrational
  free energy $\beta F_{NMA}$ in the harmonic approximation, calculated from a
  sample of 1000 minima. \textit{Left:} Src-SH3 domain. \textit{Right:} ww
  domain. An arbitrary offset was added to shift the origin of the ordinate. } 
\label{fig5}
\end{figure}

Assuming that the contribution to the integral (\ref{fdefine}) can be
approximated by local normal modes in the vicinity of the energy minimum, the
effective free energy can be written as  
\begin{eqnarray}
\beta F_{NMA}({\alpha},T)&=& \sum_{q=1}^{3N-6}\log\left(\frac{\hbar
  \omega_q(\alpha)}{k_BT} \right) = \sum_{q=1}^{3N-6}\log(\omega_q(\alpha)/\omega_q(0)) -
\sum_{q=1}^{3N-6}\log(k_BT/\hbar \omega_q(0))\nonumber\\& =& f_{NMA}({\alpha})+ \beta  F_{NMA}(0,T)\; 
\label{harmapprox} \ \ \ ,
\end{eqnarray}
where $\omega_q(0)$ are the normal mode frequencies at the ground state $\alpha=0$. 
This expression allows us to calculate a harmonic approximation to
$F_v(\alpha,T)$ by calculating the normal modes for each minimum that we sample,
and subsequently summing their different contributions according to
(\ref{harmapprox}). Figure \ref{fig5} shows the
${\alpha}$-dependent part $f_{NMA}({\alpha})$ as a function of the
energy minimum for the two examples of src-SH3 domain (left) and ww domain
(right). The minima were obtained along a single trajectory close to the folding
temperature. The distribution of the minima along the energy axis reflects the
character of the probability distribution function for the two proteins, one
being divided into two basins (src-SH3 domain), the other being a single
distribution (ww domain). 

As a first important observation, we note that the variation of $f_{NMA}({\alpha})$ with
 $e_{\alpha}$ cannot be assumed to be negligible
as assumed in (A2), according to which $F_{NMA}$ should be approximately constant in $e_{\alpha}$. Both proteins show the same trend toward a decreasing
effective free energy with increasing $e_{\alpha}$. For the src-SH3 domain, a
nonlinear dependence can be observed in the high energy range 
in agreement with
previously reported results on protein G \cite{wall}.\\A second point
to be noticed is that, for a given $e_{\alpha}$,
a distribution of values of $f_{NMA}(\alpha)$ can be found (variance on
the y-axis in figure \ref{fig5}). While such a variance (or fluctuation) could be attributed to the limited numerical
accuracy of the normal modes at high energies, this is also true for the low
energy part for which the numerical scheme provides accurate results, as can be
checked on the vanishing six lowest eigenmodes. Consequently, it appears that
the first approximation (A1), i.e. $F(\alpha,T)\approx F(e_{\alpha},T)$ does
not hold as assumed, leaving the possibility that the observed deviations in
section \ref{section3} could be caused by this simplification.

\subsection{Free energy correction}

The results of the previous subsection seem to challenge the validity of 
the ISL approach performed under the assumptions (A1) and (A2). At a first glance,
the main problem seems to arise from 
the approximation (A2) that $F_{v}(\alpha,T) $ does
not depend on the particular basin considered, which is obviously 
untrue.  On the other hand,  although there is clearly a variance associated to
$F_{v}$ for the same energy range $e_{\alpha}$ in disagreement with approximation (A1), one
can observe a general evolution of $F_{v}(\alpha,T) $ with 
$e_{\alpha}$, which suggests that approximating  $F_{v}(\alpha,T) $
by  $F_{v}(e_{\alpha},T) $ may  be acceptable.
However, we show below that, if the free energy within a basin
can be written as a sum  
\begin{eqnarray}
\beta F_v(\alpha,T) &=&f_{v}(e_{\alpha})+\beta  F_{v}(0,T) \; ,
\label{separation}
\end{eqnarray}
with $  f_{v}(e_{\alpha}=0):=0$ at the ground state, the calculation of $C_{V,IS}$ {\em can be carried out without any change,} so
that approximation (A2) appearing to be to particularly bad at a first glance
may not be the decisive one.
It should be noticed that, as discussed in the previous section, the
property (\ref{separation}) is
 verified if the motion in each basin of attraction can be
described by a combination of harmonic vibrations.\\
Starting again from (\ref{eq1}), and 
proceeding as in section \ref{ihstheory}, 
we can eliminate the part of $F_v(\alpha,T)$ that depends 
on temperature only in the
expression for the partition function and the probability distribution
function, keeping only the $\alpha$-dependent part. One gets
\begin{eqnarray}
Z(T) &=&e^{- \beta F_{v}(0,T)} 
\int\Omega_{IS}(e_{\alpha})e^{-\beta e_{\alpha}}e^{- f_{v}(e_{\alpha})}\ \ \ 
d e_{\alpha}\ \ \ \ \ , \label{newz}\\
Z_{IS}(T) &=& 
\int\Omega_{IS}(e_{\alpha})e^{-\beta e_{\alpha}}e^{- f_{v}(e_{\alpha})}
d e_{\alpha}\ \ \ . \label{newzis}
\end{eqnarray}
The principle of the calculation is to compute $Z_{IS}(T)$ form a
\textit{measurement} of the probability density function $P_{IS}(T)$
estimated from MD simulations at a given temperature $T_{MD}$. This
can be achieved through the intermediate calculation of a density of
states of inherent structures $\Omega_{IS}(e_{\alpha})$, which is
temperature independent and from which $Z_{IS}(T)$ can be obtained at
all temperatures. In equation (\ref{newzis}), we notice that the
inclusion of the term $e^{-f_{v}(e_{\alpha})}$ is equivalent in
definition an ''effective'' density of states $\Omega_{IS}(e_{\alpha})
e^{-f_{v}(e_{\alpha})}$ from which the classical thermodynamic
expression can be derived as shown below. 
 In the calculation of section \ref{ihstheory}, the density of states
 is given by equation  (\ref{eq:piszis}). This density
 $\Omega_{IS}^{(0)}$ and all other observable calculated in section
 \ref{ihstheory} will henceforth denote with and index $(0)$. In the
 new scheme including the $\alpha$-dependent part of the free energy
 in the harmonic approximation, the probability density
 $P_{IS}(e_{\alpha},T)$ becomes 
\begin{eqnarray} 
P_{IS}(e_{\alpha},T)
&=&
\frac{1}{Z_{IS}(T)}\Omega_{IS}(e_{\alpha})e^{-\beta e_{\alpha}}e^{-
  f_{v}(e_{\alpha})}\ \ \ .  
\end{eqnarray} 
yielding the inherent structure density 
\begin{eqnarray}
\Omega_{IS}(e_{\alpha})
&=& \frac{P_{IS}(e_{\alpha},T_{MD})}{p_{0}(T_{MD})}
e^{\beta_{MD} e_{\alpha}}e^{ f_{v}(e_{\alpha})}\ \ \ .
\end{eqnarray}
The latter expression shows that if the variation of $F_v(\alpha,T)$
with $e_{\alpha}$ cannot be ignored, the previously derived density of
states $\Omega_{IS}^{(0)}(e_{\alpha})$ is not the correct one. The two
are related by  
\begin{eqnarray}
\Omega_{IS}(e_{\alpha})&=&\Omega_{IS}^{(0)}(e_{\alpha})e^{
    f_{v}(e_{\alpha})}\label{relatedens} \ \ \ \ .
\end{eqnarray}
A similar result was also reported in \cite{wall} which considered the
particular case 
of a piecewise linear dependence on $e_{\alpha}$. Though the densities
differ, substituting (\ref{relatedens}) in equations (\ref{newzis}) or
(\ref{eq:piszis}), we immediately have 
$Z_{IS}(T)=Z^0_{IS}(T)$, and the inherent structure 
observables such as  $U_{IS}=\langle e_{\alpha} \rangle$ and 
$C_{V,IS}=\left(\langle
e_{\alpha}^2\rangle -\langle e_{\alpha}\rangle^2\right)/k_BT^2$ are unchanged,
 i.e. $U_{IS}=U_{IS}^{(0)}$ and $C_{V,IS}=C_{V,IS}^{(0)}$ even
 when  $f_{v} \neq 0$.\\As a consequence, using equation (\ref{newz}),
 the full free energy $F(T)$ of the protein can be written as 
 \begin{eqnarray}
F(T)&=&-k_BT\log(Z(T))=-k_BT\log(Z_{IS}(T))+  F_{v}(0,T)\nonumber
\\ &=&-k_BT\log(Z_{IS}^{(0)}(T))+ F_{v}(0,T)=F_{IS}^{(0)} + F_{v}(0,T)\ \ \ \ .
\end{eqnarray}
Therefore, in the ISL formalism, taking into account the variation of
$F_v(\alpha,T)$ as in equation (\ref{separation}) does not alter the
free energy and cannot be expected to be at the origin of the
quantitative differences between $C_V$ derived from the ISL formalism
and the full numerical results presented in section \ref{section3}. We
conclude that the origin of these discrepancies is likely to be found
in the non-separability between the $\alpha$ and the $T$ dependency
within the basins. Such an non-separability can be expected as soon as
the anharmonicity of the different basins is taken into account. This
is certainly relevant for proteins, in particular as the denaturation
involves frequent transitions between basins of different shape and
volume associated to the semi-rigid folded and the highly flexible
unfolded state.\\When $F_v(\alpha,T)$ cannot be separated into
$\alpha$- and $T$-dependent contributions to simplify the calculation
of the partition function, the remaining possible approximation that
can tackle the computational difficulty would be the saddle point
approximation of the free energy \cite{stillinger}. This approximation
is acceptable in the thermodynamic limit for large systems, but cannot
be justified in the present problem as the number of particles
involved is still small and the interactions between particles are
heterogeneous. While the harmonic approximation seems to be invalid
for the present problem which involves large conformational changes
due to the denaturation transition, results on super-cooled liquids
\cite{sciortino} indicate that the correction of the heterogeneity of
the basins at low temperatures is small and the decoupling
approximation of vibrational and inherent structure contributions
appears to be possible at least in these temperature regimes. 

\subsection{Effect of limited sampling efficiency}

The main practical difficulty of the ISL method comes from the need to
properly sample all the inherent structures in order to get a meaningful
density of states $\Omega_{IS}(e_{\alpha})$ in all energy ranges.  In the
present scheme of inherent structure sampling, a single temperature is
selected to simulate the dynamics of the protein for a finite period of
time. The choice of a temperature close to the folding temperature is natural
as the protein samples both the folded and
the unfolded configurations at this temperature. There are however two
questions that have to be 
answered: \textit{i)} How long should we follow a protein MD trajectory to get a
sufficient sampling? \textit{ii)} Is it possible to combine data from
simulations 
at a few different temperatures instead of keeping $T$ fixed?\\Let us
first analyze  
 the effect of the sampling time.  It should be chosen
long enough to cover the slowest intrinsic timescale of the system, and
 various trajectories with different initial conditions or
realizations of the thermostat should be used to ensure that the order
of events does 
not alter the shape of the distribution.  An estimate of the time range that
sampling must cover is provided by the folding/unfolding time of the protein,
to guarantee that the protein explores both configurational
subspaces. A possible check 
of the choice of the simulation time is to compute the specific heat with an
increasing number of samples, and stop when the improvement brought by
additional samples is negligible.  Still, as computer time is limited, it
cannot be ensured that all relevant states are sufficiently well sampled to
yield a converged probability density. In particular, the high energy minima
are sampled only with low probability, such that the high energy cut-off in
the density of states is likely to be underestimated in finite time
sampling. In this section, we analyze the impact of this cutoff on a model
density to see how the inherent structure specific heat is possibly affected.

\begin{figure}
\centering
\includegraphics[width=8.0cm]{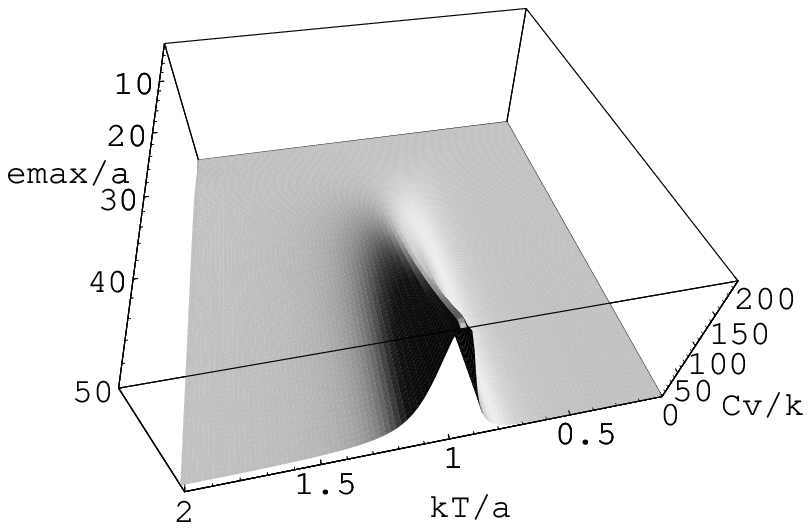}
\includegraphics[width=8.0cm]{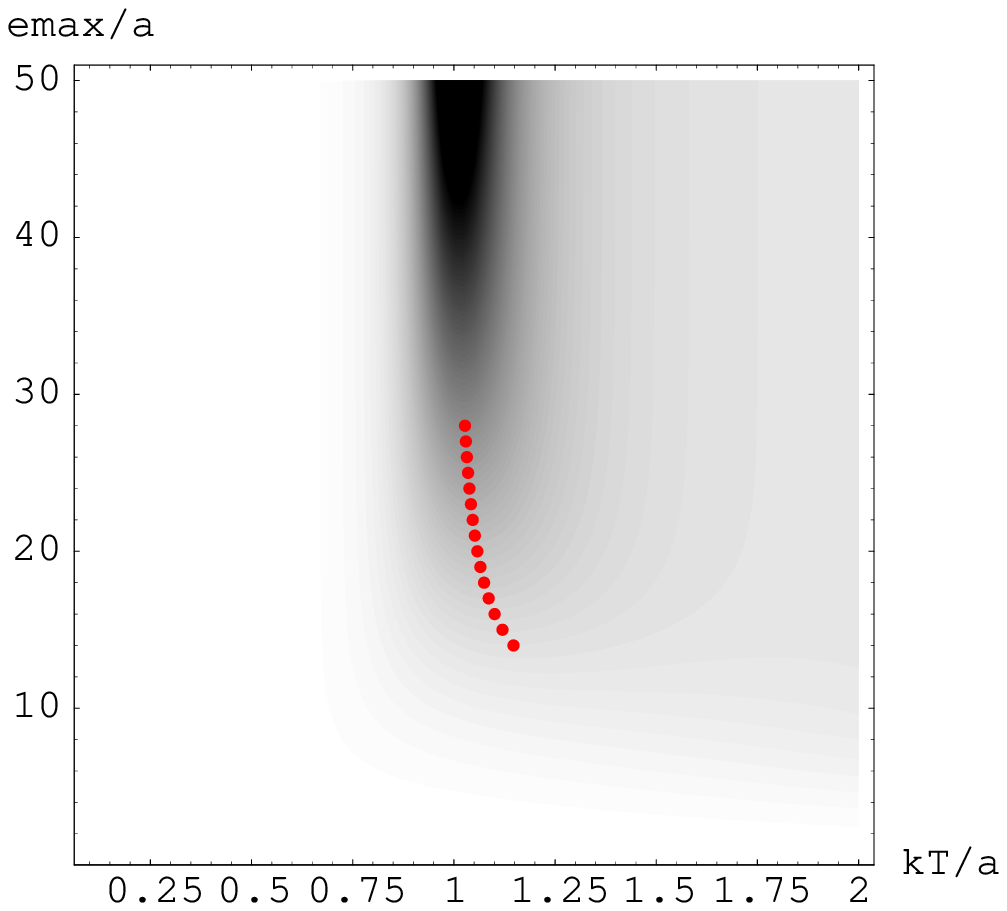}
\caption{\textit{Left:} The specific heat $C_{V,IS}$ as a function of
  temperature 
  and the cutoff parameter $e_{max}$. \textit{Right: } Location of the maxima
  of $C_{V,IS}$ for different values of the cutoff $e_{max}$.} 
\label{fig6}
\end{figure}

To analyze the effect of the sampling independently of a particular
case, let us assume a ''model'' inherent structure density of states taken as a
single exponential  
\begin{equation*}
\Omega_{IS}^{(0)}(e_{\alpha})= \begin{cases}
 e^{e_{\alpha}/a}  & 0\le e_{\alpha} \le e_{max} \\
  0 &  e_{max} < e_{\alpha}
\end{cases}  \ \ \ ,
\end{equation*}
similar to the shape of the densities that can be found in limited ranges of
energies for the numerical results in section \ref{section3}. The partition
function than can be readily calculated as 
\begin{eqnarray}
  Z_{IS}^{(0)}&=&\int_{0}^{e_{max}}
  de_{\alpha}\  \Omega_{IS}(e_{\alpha})\ e^{-\beta
    e_{\alpha}}=\frac{e^{e_{max}(a^{-1}-\beta)}-1}{a^{-1}-\beta}\ \ \ . 
\end{eqnarray}
Likewise, we can calculate the first two moments $\langle e_{\alpha}\rangle$
and $\langle e_{\alpha}^2\rangle$ to find the specific heat as a function of
the temperature, the parameter $a$ and cutoff in energy $e_{max}$ 
\begin{eqnarray}
C_{V,IS}^{(0)}(T;a,e_{max})&=&\frac{\langle e_{\alpha}^2\rangle-\langle
  e_{\alpha}\rangle^2}{k_B T^2} \ \ \ ,
\end{eqnarray}
and analyze the result graphically as a function of the energy cutoff
$e_{max}$ in figure \ref{fig6}.  
As can be inferred from left-hand side of figure \ref{fig6},
$\max_{T}\left[C_{V,IS}(T;a,e_{max})\right]$ 
 increases with higher cut-off $e_{max}$. Using
symbolical computation \cite{Mathematica}, 
we can further inspect the
result to find the local maxima of the the specific heat for fixed cutoff
$e_{max}$. On the right-hand side of figure \ref{fig6}, we observe
that a lower cutoff in 
the density of states shifts the maximum of the specific heat towards higher
temperatures. In addition to the shift, the curve becomes broader and the
value at the maximum decreases. This situation is similar to the physical
scenario when protein folding is altered by confinement (see
e.g. \cite{depablo}). The high energy states disappear from the density of
states as the system is prohibited to explore these by external forces. For
the present purpose of the inherent structure analysis, although derived for a
highly idealized model density, the results indicate that insufficient
sampling at high energies can significantly alter the global shape of the
transition. We have checked that these conclusions remain unchanged for a
piecewise constant density of states. For instance, for the ww domain,
one finds $e_{max}/a\approx 60$ which is higher than the range of
values for which a shift of the maximum can be expected from figure
\ref{fig6}. Consequently, the origin of this shift cannot be
attributed to inefficient sampling in the high energy range.\\Because
the high energy minima are  
less frequently visited, it is tempting to
try to sample the minima from a high temperature molecular dynamics trajectory. 
On the other hand, as the method relies on the probability 
to occupy the ground state which determines $1/Z_{IS}(T)$, it is also
necessary to properly sample the ground state, i.e. to select a simulation
temperature which is below $T_f$. To reconcile these two exclusive 
conditions, one solution is to combine results sampled at two different
temperatures to calculate $\Omega_{IS}(e_{\alpha})$, which should be
temperature independent. This is possible because, according to
(\ref{eq:pz}) 
\begin{equation}
\label{eq:zratio}
  \dfrac{P_{IS}(e_{\alpha},T_1)}{P_{IS}(e_{\alpha},T_2)}
= \dfrac{Z_{IS}(T_2)}{Z_{IS}(T_1)} e^{-(\beta_1-\beta_2)e_{\alpha}} \; \ \ \ ,
\end{equation}
with $\beta_{1,2} = 1/(k_B T_{1,2})$, so that the ratio of
${Z_{IS}(T_2)}/{Z_{IS}(T_1)}$ can be calculated from the 
probabilities to occupy a basin at temperatures $T_1$ and $T_2$.  A molecular
dynamics trajectory obtained at a temperature $T_1 < T_f$ can be used to
determine $Z_{IS}(T_1)$ from the probability to occupy the ground state, and
subsequently a second simulation at a higher temperature $T_2$ can
sample high energy 
basins more efficiently. For all basins which are properly sampled in both
molecular dynamics runs, the ratio ${Z_{IS}(T_2)}/{Z_{IS}(T_1)}$ can be
evaluated with (\ref{eq:zratio}). Although it should not depend on the
particular basin that was used for its calculation, this ratio actually
fluctuates around a mean value which can be used to determine ${Z_{IS}(T_2)}$
from ${Z_{IS}(T_1)}$. Then (\ref{eq:pz}), applied at the higher
temperature $T_2$, can be used to compute $\Omega_{IS}(e_{\alpha})$ in the
high energy range. Moreover, in the intermediate energy range, the basins are
properly sampled by the two trajectories at temperatures $T_1$ and $T_2$,
which gives two ways to evaluate $\Omega_{IS}(e_{\alpha})$ for those basins,
and thus provides a way to check the consistency of the method.
\begin{figure}[H]
\centering
\includegraphics[width=7.8cm]{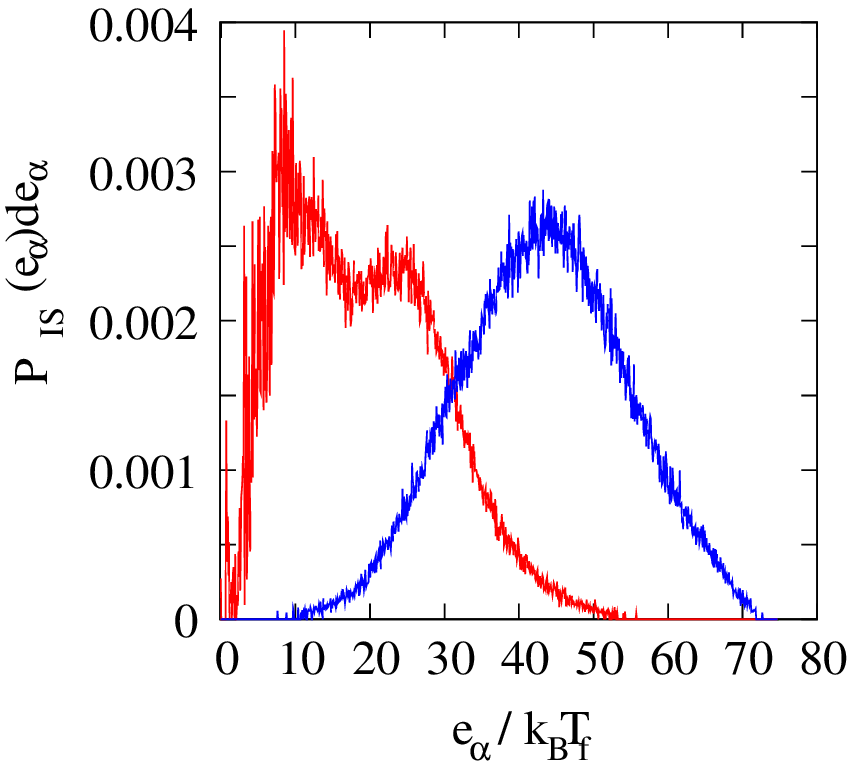} 
\includegraphics[width=7.05cm]{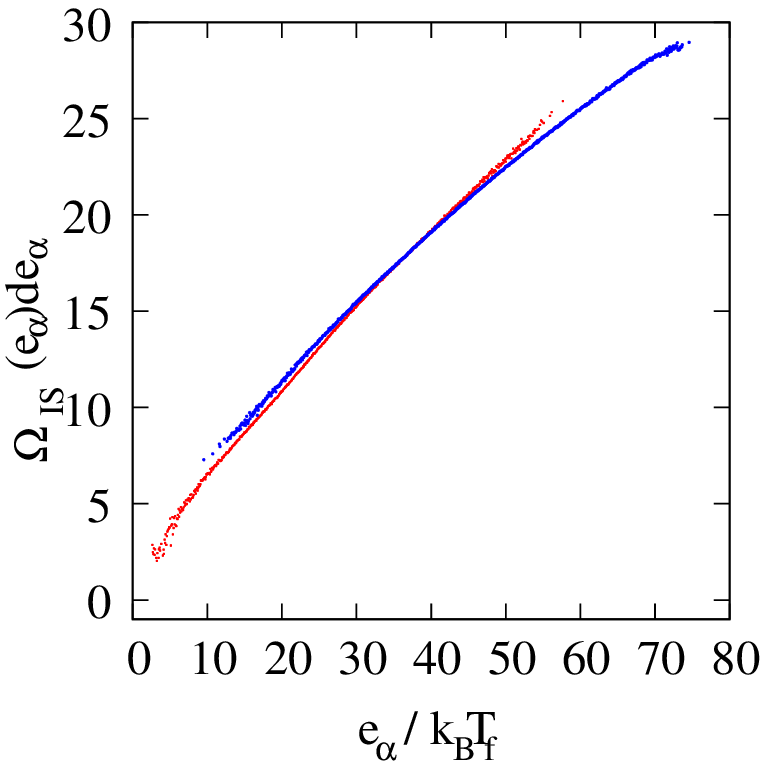} 
\caption {
Results from the sampling of inherent structures of the ww domain
    at two different temperatures $T_1=1.03\ T_f$ ($\approx 79000$ minima, red)
and $T_2=1.41\ T_f$ ($\approx 18000$ minima, blue) 
\textit{Left:} Probability of occupation of the inherent structures
versus $e_{\alpha}$. \textit{Right: } Density of inherent states
energy (in logarithmic scale),  
calculated from the simulation at
$T_1$ (red) and calculated at $T_2$ using the ratio
 ${Z_{IS}(T_2)}/{Z_{IS}(T_1)}$.}
  \label{fig:2T}
\end{figure}
Figure \ref{fig:2T} shows the results for the ww domain from different molecular
dynamics simulations, simulated respectively at $T_1/T_f=1.03$ and
$T_2/T_f=1.41$.  
 For the simulations at $T_2$, 
the ground state is not sampled in the finite interval of time of the
simulations, 
but $\Omega_{IS}$ can nevertheless be obtained through the evaluation of
$Z_{IS}(T_2)$ deduced from $Z_{IS}(T_1) = 1/p_0(T_1)$ for all values of
$e_{\alpha}$ for which the histograms of the left-hand side of figure
\ref{fig:2T} overlap. 
The right-hand side of figure \ref{fig:2T} shows that the values of
the density of inherent 
state energies computed from data at $T_1$ and $T_2$ are in rather
good agreement in
the whole energy range where the basins are sampled at the two
temperatures. There is however a discrepancy between the two results,
with a systematic deviation towards lower values of $\Omega_{IS}(e_{\alpha})$
for high inherent state energies when the density of states is calculated with
data sampled at high temperatures. This is counterintuitive as one
could tend to believe that sampling the basins at low temperatures
would, on the contrary, 
underestimate the density of basins at high energy. This systematic deviation,
which has been observed in all our calculations could point out a limitation
of the ISL method as presently applied, i.e.\ with a calculation which
is only valid 
if $\beta F_v(\alpha,T)$ is the sum of a term that depends on $e_{\alpha}$ and
a term that depends on temperature (see (\ref{separation})~). 
When the specific heat is calculated with the value of
$\Omega_{IS}(e_{\alpha})$  
extended in the high energies range with this method, the function is
still very close 
to the value obtained with only the data at temperature $T_1$ and the
agreement with the numerical value of $C_V(T)$ is not improved. This shows
that the discrepancy between the exact results and those deduced from the ISL
approach are not due to technical difficulties such as an insufficient sampling
in some energy range, but that they are rather inherent to the method
itself together 
with its assumptions.

\section{Discussion}
\label{section5}

We applied the inherent structure landscape (ISL) approach to four different
proteins of varying size and secondary structure elements using a
coarse-grained off-lattice protein model, and calculated their inherent
structure density of states. Using these densities, we derived the specific
heat from the reduced inherent structure thermodynamics, and compared it to
the value obtained from equilibrium molecular dynamics as a function of
temperature. Our results show that the ISL approach can correctly capture the
shape of the temperature dependence of the specific heat, including some
characteristic features such as the hump observed at $T \approx 0.4 T_f$ for
the ww domain. This is remarkable since the result is deduced from
molecular dynamics simulations at a single temperature, close to $T_f$, and
nevertheless predicts the main features of the specific heat in a large
temperature range, including low temperatures for which very long simulations
would be necessary to reach exhaustive sampling of the phase space. This shows
that {\em many features of protein thermodynamics are encoded in
  the inherent structure landscape.} Still, the approach is not perfect
as we observed quantitative differences between $C_{V,IS}(T)$ and
$C_V(T)$ which are particularly significant for small protein domains.  
The deviations show a systematic trend,
the specific heat being underestimated below $T_f$ and overestimated above.
This lead us to reexamine the approximations that enter the
construction of the reduced thermodynamics from inherent structures for the
model that we considered. 
The first approximation assumes that the correction to
the density of states due to the structure of the basin associated to a
minimum $f_{v}(\alpha)$ depends on the energy level of the minimum
only, and not of the individual minimum. An evaluation of
$f_{v}(\alpha)$ using a harmonic approximation  based on local
normal modes (section \ref{NMA}) shows that this is only approximately correct.
For a given $e_{\alpha}$, the values of $ f_{v}(\alpha)$ are actually
distributed around an average value, with fluctuations that grow for higher
values of $e_{\alpha}$ and that appear to be larger for small
proteins. This could explain some 
of the discrepancies between the ISL results and the equilibrium data.
Moreover, the calculation of $ f_{v}(\alpha)$ indicates that the
second assumption that the correction can be considered to be
$\alpha$-independent is certainly not valid. However, we showed that if $\beta
F_v(\alpha)$ splits into a temperature-dependent and an $\alpha$-dependent
part, which is the case in the harmonic approximation, most of the
thermodynamic results deduced from a direct application of the ISL approach
are not affected. This is true in particular for the the specific heat.
In view
of the persisting quantitative differences between reduced inherent structure
and equilibrium thermodynamics, we therefore conclude that the correction of
the free energy in term of a harmonic approximation is not sufficient.
It is likely that the 
nonlinear terms in the free energy associated to a basin cannot be ignored,
and play a significant role. This is not surprising because,
especially in the high temperature range, the protein fluctuates by exploring
many basins, and consequently cannot be assumed to be well described by a
 harmonic approximation.\\
In future studies, it would be useful to analyze the role of the structure of
the full basin on the thermodynamic results beyond the approximation by local
normal modes around the minima.  This is a true challenge owing to the
complexity of the energy landscape. A starting point for such a study might be
the examination of the distribution of first rank saddle points associated to
the different minima on the potential energy surface. A second aspect which is
suggested by the present work is to apply the ISL approach for protein folding
in the context of more complex energy landscapes that arise in more realistic
potential energy functions. The results on the small proteins analyzed in this
study show that the global separation of the probability density into two
basins associated to folded and unfolded states is not a necessary requirement
to construct the reduced thermodynamics. It appears therefore likely that the
formalism remains useful in cases where the energy landscape is less biased
towards the ground state than in the G\=o-model. An application of the ISL
method to other protein models therefore appears to be desirable and
promising.

\bigskip

\textbf{Acknowledgements:\ }We thank the anonymous referee for valuable 
suggestions. J-G.H. acknowledges financial support from the
Coll\`ege Doctoral Franco-Japonais and the R\'egion Rh\^one-Alpes. The
simulations were partially performed at the P\^ole Scientifique de
Mod\'elisation Num\'erique (PSMN) in Lyon.

\appendix

\section{Model Hamiltonian and parameters}
\label{ap1}

In this work, we analyze the properties of an off-lattice G\=o-type
\cite{nakagawa1,nakagawa2} in which the smallest building unit is a
single amino acid. Effective interactions between amino acids are
based on the reference positions of the $C_{\alpha}$-carbons of each
residue. These interactions are ''color-blind'' in the sense that they
do not distinguish between the type of amino acids. The potential
energy of the system comprises five terms: 
\begin{eqnarray}
  V&=&\sum_{i=1}^{N-1} \frac{1}{2} K_h (d_i-d_{i0})^2\ +\
  \sum_{i=1}^{N-2}\frac{1}{2} K_b (\theta_i-\theta_{i0})^2 \nonumber
  \\
  & &+\ \sum_{i>j-3}^{native} \epsilon
  \left[5\left(\frac{r_{0ij}}{r_{ij}}\right)^{12}-6\left(\frac{r_{0ij}}{r_{ij}}
\right)^{10} 
    \right] + 
  \sum_{i>j-3}^{non-native} \epsilon
  \left(\frac{C}{r_{ij}}\right)^{12}\nonumber\\ 
  & &+ \ 
  \sum_{i=1}^{N-3}K_d
    \left(1-cos\left(2\phi_i-\frac{\pi}{2}\right)\right) 
\end{eqnarray}  
Here, $r_{ij}$ denotes the Euclidean distance between residues
$i,j$. The $3N-6$ degrees of freedom of the system are most
conveniently expressed via internal coordinates: $N-1$ Euclidean bond
distances $d_i$ along the backbone, $N-2$ bond angles formed between
two consecutive bond directions, and $N-3$ dihedral angles measured
between the normal vectors of planes spanned by atoms $(i,i+1,i+2)$
and $(i+1,i+2,i+3$). Bond distance and bond angle interactions are
modelled through harmonic forces with coupling strengths $K_h$ and
$K_a$ respectively. The zero indices indicate that the quantities
(angles, distances) are evaluated in the reference state, i.e., the
position from the NMR/crystallographic structure. The dihedral angle
potential does not assume a minimum in the reference position defined
by the experimentally resolved structure: it favors angles close to
$\pi/4$ and $3\pi/4$ irrespective of the secondary structure element
(helix, sheet, turn) the amino acid belongs to. While such values can
be found in $\alpha$-helices, they statistically do not appear largely
in other secondary structure elements. As a consequence, the reference
state defined by an NMR or crystallographic structure can give rise to
a competition between the helix for which the reference state is close
to its minimum energy, and other parts which experience forces due to
the angular constraint. The constraint was introduced as a source of
additional ''frustration'' affecting the dynamics and thermodynamics
of the model towards a more realistic representation
\cite{nakagawa1}. Attractive non-bonded native interactions are
modelled by steep (12,10)-Lennard-Jones potentials accounting for
non-local contacts between residues due to side chains. The database
of contacts is established as described in section \ref{section2}. In
addition, non-native repulsive interactions are added to those
residues which do not form a contact and which lie at least four
residues apart. The dimensionless parameters used in this study are
\cite{nakagawa1}: $K_b=200.0$, $K_a=40.0$, $K_d=0.3$, $\epsilon=0.18$,
$C=4.0$.

\end{document}